\documentstyle[12pt]{article}

\def\L{{\cal L}}
\def\ka{{\kappa}}
\def\la{{\lambda}}
\def\ba{{\beta}}
\def\ga{{\gamma}}

\newcommand{\be}{\begin{eqnarray}}
\newcommand{\en}{\end{eqnarray}}

\begin{document}
\begin{titlepage}
\begin{flushright}
EFI 95-47 \\
MPI-Ph/95-81  \\
\end{flushright}

\begin{center}
\vskip 0.3truein

{\bf {Reduction of Coupling Parameters}}
\footnote{Plenary talk presented at the XVIIIth International
   Workshop on High Energy Physics and Field Theory,
   Moscow-Protvino, June 1995. To be published in the Proceedings.}
\vskip0.5truein

{Reinhard Oehme}
\vskip0.2truein

{\it Enrico Fermi Institute and Department of Physics}

{\it University of Chicago, Chicago, Illinois, 60637, USA} \footnote{Permanent
Address}

{\it and}

{\it Max-Planck-Institut f\"{u}r Physik}

{\it - Werner-Heisenberg-Institut -}

{\it 80805 Munich, Germany}
\end{center}
\vskip0.2truein
\centerline{\bf Abstract}
\vskip0.13truein

The general theory of the reduction in the number of coupling
parameters is discussed. The method involves renormalization
group invariant relations between couplings. It is more
general than the imposition of symmetries. There are reduced
theories with no known symmetry. The reduction scheme is finding
many applications. Discussed in some detail are
the construction of gauge theories with ``minimal'' coupling
for Yang-Mills and matter fields, and the Gauge-Yukawa
Unification within N=1 supersymmetric GUT's.

\end{titlepage}
\newpage
\baselineskip 20 pt
\pagestyle{plain}
\setcounter{equation}{0}

\centerline{\bf I. INTRODUCTION}
\vskip0.2truein

In recent years, the method of the reduction in the number of
coupling parameters of renormalizable field theories
\cite{OZS,NPC} has found
many applications, theoretical and phenomenological.  In this talk,
we will briefly review \cite{CON} the reduction scheme, and report about
some of the applications.

Lagrangians of quantum field theories can be constrained by
renormalizability requirements.  Generally, this leaves a large
number of independent coupling parameters.  We consider the
possibility to reduce this number by imposing relations between the
dimensionless couplings such that renormalizability is preserved,
and that the relations are independent of the normalization point.
We require that the original, as well as the
reduced theories satisfy the corresponding renormalization
group equations. The resulting
set of differential equations, the reduction equations, are necessary
and sufficient for the renormalization group invariance of the reduced
theory.

A standard method for the reduction in the number of couplings
is the imposition of symmetries.  The reduction method described
above is more general.  It includes possible solutions reflecting
symmetries, but also provides other reductions which have no known
connections with any symmetry.

\noindent In the following, we list some of the
applications of the reduction method:

* Construction of gauge theories with ``minimal'' coupling of
Yang-Mills and matter fields \cite{OSZ}.

* Classification of renormalizable theories with a single coupling
parameter \cite{OZS}.

* Proof of conformal invariance (finiteness)  for $N=1$ SUSY  gauge
theories with vanishing lowest order $\beta $-function
on the basis of one-loop information \cite{LPS}.

* Reduction of the infinite number of coupling parameters appearing
in the light-cone quantization method \cite{WPL}.

* Gauge-Yukawa unifications within the framework of SUSY GUT's.
Successful calculations of top-quark and bottom-quark masses within
this framework \cite{KMZ}.

* Applications of reduction to the standard model
(non-SUSY) give values for the top-quark mass which are too small,
indicating the need for more matter fields \cite{KSZ,RON}.

* Universal soft breaking of supersymmetry. Renormalization group
invariant relations between soft SUSY-breaking parameters \cite{DRT}.

There are many other applications. The reduction of
Yukawa couplings may also be of interest in connection
with duality and IR-fixed points in $N = 1$ SUSY gauge theories.

\vskip0.2truein
\centerline{\bf 2.  REDUCTION SCHEME}
\vskip0.1truein

Let us consider a renormalizable theory with $n+1$ real,
 independent
dimensionless parameter, $\la, \la_1,\ldots, \la_n$.
 We call
$\la$ the ``primary coupling'', and consider the
reduction to one
coupling parameter by the requirement
\be
\la_k = \la_k (\la), ~~~k = 1,\ldots,n ~.
\label{2.1}
\en
The Green's functions of the reduced theory should then
satisfy the
renormalization group equation
\be
\left( \ka^2 \frac{\partial}{\partial \ka^2} + \beta_{\la} (\la)
\frac{\partial}{\partial\la} + \gamma_G(\la) \right)
G(k_i,\ka_2,\la) ~ = ~ 0 ~~,
\label{2.2}
\en
where $G$ is defined in terms of the multi-parameter Green's
functions by
\be
G(k_i,\ka^2,\la) = G \left( k_i, \ka^2, \la, \la_1 (\la),\ldots,
\la_n (\la)\right),
\label{2.3}
\en
with corresponding expressions for the coefficients:
\begin{eqnarray}
\ba_{\la}(\la) ~&=&~ \ba_{\la} (\la ,\la_1(\la) , \ldots ,
\la_n (\la) )~,    \cr
\ga_{G}(\la) ~&=&~ \ga_{G} (\la ,\la_1(\la) , \ldots ,
\la_n (\la) )  ~~.
\label{2.3a}
\end{eqnarray}
The Green's functions of the original theory satisfy
the equations
\be
\left( \ka^2 \frac{\partial}{\partial \ka^2} + \beta_\la
\frac{\partial}{\partial\la} + \sum^n_{k=1} \beta_k
\frac{\partial}{\partial{\lambda_k}} + \gamma_G\right) G~=~0~,
\label{2.4}
\en
where $G=G (k_i,\ka^2,\la,\la_1,\ldots,\la_n)~$ ,
$\beta_\la = \beta_\la (\la,\la_1,\ldots,\la_n)~$, etc. .
By substitution,
we find then the {\it reduction equations} :
\be
\beta_\la (\la) \frac{d\lambda_k (\la)}{d\la} = \beta_k (\la)~,
{}~~~k=1,\ldots,n
\label{2.5}
\en
as necessary and sufficient conditions for the validity
of the renormalization group equations (\ref{2.2} )
for the reduced theory.

We consider here massless theories, or theories
with mass independent renormalization schemes, so that no mass
parameters occur in the coefficient functions of the renormalizaton
group equation. The perturbative renormalizability of the
multi-parameter theory implies then that the $\beta$-functions
have asymptotic power series expansions at $\la=\la_k = 0$.
For the models considered, these expansions can be
written the form
\be
\beta_\la (\la,\la_1,\ldots,\la_n) = \beta_{\la 0} \la^2 +
(\beta_{\la 1} \la^3 +
\beta_{\la 1,k} \la_k \la^2 + \beta_{\la 1,k k'} \la_k \la_{k'}
\la ) + \cdots ~,
\label{2.6}
\en
and
\be
\beta_k (\la,\la_1,\ldots,\la_n) = (c^{(0)}_k \la^2 +
c^{(0)}_{k,k'} \la_{k'} \la + c^{(0)}_{k,k' k''}\la_{k'}\la_{k''}) +
\cdots ~.
\label{2.7}
\en
We note that $\beta_\la$ vanishes for $\la \rightarrow 0$.

Since $\beta_\la (\la)$ generally will have a zero at $\la=0$, this is a
singular point of the reduction equations and the usual existence
theorems are not applicable here.  Without further constraints, we may
have general solutions with as many free parameters as there are couplings.
These are not of direct use for the reduction, but will be considered
later in connection with small coupling treated as corrections.  For the
actual reduction, we make the assumption that the Green's functions
of the reduced theory have power series expansion in $\la$.  To this
end, we require that the couplings $\la_k (\la)$ are given by
asymptotic power series for $\la\rightarrow 0$.

\vskip0.2truein
\centerline{\bf 3.  SPECIAL SOLUTIONS}
\vskip0.1truein

With all couplings $\la,\la_k$ vanishing on the weak coupling limit
of the original theory, we assume that
\be
\la_k (\la) = \la f_k (\la),~~~ k=1,\ldots,n ~,
\label{3.1}
\en
with bounded functions $f_k (\la)$ in some interval $0\le\la\le\la_0$.
It is then useful to write the reduction equations (\ref {2.5})
in terms of $f_k (\la)$:
\be
\beta_\la \left( \la \frac{d f_k}{d\la} + f_k\right) = \beta_k ~,
\label{3.2}
\en
with $f_k (\la)$ given by the power series expansions
\be
f_k (\la) = f^0_k + \sum^\infty_{m=1} \chi^{(m)}_k \la^m ~.
\label{3.3}
\en
It is convenient in the following to use the notation
\be
\beta_\la (\la) = \beta_\la (\la,\la f_i (\la)) = \sum^\infty_{n=0}
\beta_{\la n} (f)\la^{n+1} ~,
\label{3.4}
\en
\be
\beta_k (\la) = \beta_k (\la,\la f_i (\la)) = \sum^\infty_{n=0}
\beta^{(n)}_k (f) \la^{n+2} ~.
\label{3.5}
\en
Now we substitute the expansion (\ref{3.3}) into the
reduction equations (\ref{3.2}).  In lowest order, using the notation of
(\ref{3.4}) and (\ref{3.5}), we obtain the one--loop relations
\cite{OZS,NPC} :
\be
\beta_{\la 0} (f^0) f^0_k - \beta^{(0)}_k (f^0) ~=~0 ~ ,
\label{3.6}
\en
or, in explicit form, with the coefficients as defined in (\ref{2.7}):
\be
c^{(0)}_k + \left( c^{(0)}_{k,k'} - \beta_{\la 0}
\delta_{kk'}\right) f^0_{k'} + c^{(0)}_{k,
k^\prime k^{\prime\prime}} f^0_{k^\prime} f^0_{k^{\prime\prime}}~ =~ 0
\label{3.7}
\en
Here, and in the following, appropriate summation over equal
indices is understood.

The equations (\ref{3.6}) are the fundamental relations for special
reductions.  One loop criteria also decide whether the higher order
coefficients are determined by the reduction equations.  Up to
$m+1$ loops, we have the equations:
\be
\left(M_{kk^\prime} (f^0) - m \beta_{\la 0} \delta_{kk^\prime}\right)
\chi^{(m)}_{k^\prime} = \left(\beta_{\la m} (f^0) f^0_k -
\beta^{(m)}_k (f^0)\right) + X^{(m)}_k ~,
\label{3.8}
\en
where $m=1,2,\ldots,~~~ k=1,\ldots, n  ~$.
The matrix $M(f^0)$ is given by
\be
M_{kk^\prime} (f^0) = c^{(0)}_{k,k^\prime} + 2 c^{(0)}_{k,k^\prime
k^{\prime\prime}} f^0_{k^{\prime\prime}} - \delta_{kk^\prime}
\beta_{\la 0}~.
\label{3.9}
\en
The rest term $X^{(m)}$ depends only upon the coefficients
$\chi^{(1)},\ldots,\chi^{(m-1)}$, and upon the $\beta$--function
coefficients in (\ref{3.4}) and (\ref{3.5}), evaluated at
$f_k = f^0_k$, for order $m-1$ and lower.  They vanish for
$\chi^{(1)} = \ldots = \chi^{(m-1)} = 0$.

We see that the {\em one--loop} criteria
\be
det \left( M_{kk^\prime}(f^0)- m \beta_{\la 0}\delta_{kk^\prime}\right)
 \not = 0
   ~~ for ~~m=1,2,\ldots
\label{3.10}
\en
are sufficient to insure that all coefficients $\chi^{(m)}$ in the
expansion (\ref{3.3}) are determined.  Then the reduced theory has a
renormalized power series expansion in $\la$.  All possible solutions
of this kind are determined by the one--loop equation (\ref{3.6}) for
$f^0_k$.

If the conditions (\ref{3.10}) are satisfied, and all coefficients
$\chi^{(m)}$ are determined, we can use regular reparametrization
transformations in order to remove all higher terms in the expansion
(\ref{3.3}).  The reparametrization transformations are of the form
\begin{eqnarray}
   \la^\prime &=& \la^\prime (\la,\la_1,\ldots,\la_n) = \la + a^{(20)}
\la^2 + a_k^{(11)} \la_k \la + \cdots  ~~, \cr
\la^\prime_k  &=&  \la^\prime_k
(\la,\la_1,\ldots,\la_n) = \la_k +
b^{(20)}_{kk^\prime k^{\prime\prime}} \la_{k^\prime}
\la_{k^{\prime\prime}}
+ b^{(11)}_{kk^\prime} \la_{k^\prime} \la + \cdots ~.
\label{3.11}
\end{eqnarray}
They leave invariant the one-loop quantities
\be
f^0_k,~ \beta_{\la 0} (f^0),~ \beta_k^{(0)} (f^0),~
M_{kk^\prime} (f^0) ~,
\label{3.11a}
\en
but not the coefficients $\chi^{(m)}$, which can be transformed to zero if
they are uniquely determined by (\ref{3.8}).  Then we have a frame
in coupling parameter space, where
the special solutions are of the form
\be
 \la_k (\la) = \la f^0_k ~ ,
\label{3.11b}
\en
with the quantities
$f^0_k $ being given exactly as solutions of the one-loop reduction equations
(\ref{3.6}).  In many (but not all)
cases, their exist only a few solutions $f^0_k$, and hence
corresponding reduced
theories with power series expansions. In particular, this is the
case after the imposition of simple physical requirement, like
positivity etc..

In the special cases where $f^0_k = 0$, and $\chi^{(m)}_k = 0$ for
$m<N$, but $\chi^{(N)}_k \not= 0$, we find that
\be
f_k (\la) = \chi_k^{(N)} \la^N,~~~ N\ge 1 ~
\label{3.11c}
\en
after appropriate reparametrization.

Let us now consider the situation where
 the matrix $\beta^{-1}_{\la 0} M_{kk^\prime} (f^0)$ has
one or more positive integer eigenvalues, so that the determinant
(\ref{3.10}) vanishes.  Suppose there is a positive eigenvalue
for $m=N\ge 1$ and $\beta_{\la 0} \not= 0$.  Then we find that the
asymptotic power series must be supplemented by terms of the form $\la^m
(lg \la)^p$, with $ m\ge N$ and $ 1 < p < \sigma (N)$ \cite{CON,RON}.
After reparametrization, we obtain then an expansion of the form
\be
f_k(\la ) = f^0_k + \chi^{(N,1)}_k \la^N \lg \la +
 \chi_k^{(N)} \la^N + \ldots ~~,
\label{3.12}
\en
where all parameters are determined except the vector $\chi^{(N)}_k$,
which contains $r$ free parameters if the eigenvalue has
$r$-fold degeneracy.
We have a well defined,
``renormalized'' theory with logarithmic terms in the asymptotic
expansion for $\la \rightarrow 0$.

We will discuss application in a later section.  Here we mention only the most
simple example of special reduction, a model with Yukawa and quartic
coupling:
\be
\L_{int} = i \sqrt{\la} ~ \overline{\psi } \gamma_5 \psi \phi ~-~
\frac{\la_1}{4!} \phi^4 ~.
\label{3.13}
\en
With $\la$ as the primary coupling, the reduction results in the
solutions $\la_{1\pm} = \la f^0_{1\pm}$ with $f^0_{1\pm} = \frac13 (1\pm
\sqrt{145})$.  The matrix is $\beta_{\la 0}^{-1} M_{11} (f^0) =
\pm\sqrt{145}/5$, so that there is no positive eigenvector.  Since
$\la_1$ should be positive, the unique, relevant solution is $\la_1 = \la
\frac13 (1 + \sqrt{145})$.

\vskip0.2truein
\centerline{\bf 4.  GENERAL SOLUTIONS}
\vskip0.1truein

For systems with non--vanishing determinant (\ref{3.10}) for all
$m=1,2,\ldots$, we may have general solutions of the reduction equations
which approach the special solutions $\la_k (\la) = \la f^0_k$ for
$\la\rightarrow 0$.  For simplicity, and as a characteristic example, let
us assume that $\beta_{\la 0}\not= 0$, and that $\beta^{-1}_{\la 0}
M(f^0)$ has one positive, non--integer eigenvalue $
\eta > 0$, all other
eigenvalues being negative.  Then we find general solutions of the form
\be
f_k (\la) = f^0_k + \sum_{a,b} \chi_k^{(a\eta + b)} \la^{a\eta +
b} + \sum_m \chi^{(m)}_k \la^m
\label{4.1}
\en
with $a=1,2,\ldots~$, $b=0,1,\ldots ~$, $a\eta + b = ~$non-integer.
After reparametrization, powers with $m<\eta$ are removed, and we have $f_k
(\la) = f^0_k + \chi^{(\eta)}_k \la^\eta + \ldots$ ,
which is of particular interest
in situations where $f^0_k = 0$, and where the corresponding couplings
can be considered as small corrections. This will be discussed later.

In (\ref{4.1}), all coefficients are determined except
$\chi^{(\eta)}_k $, which contains
$r$ arbitrary parameters if the eigenvalue has $r$--fold
degeneracy. With
\be
 \ba^{-1}_{\la 0} M_{kk'}(f^0) \xi^{(i)}_{k'} = \eta \xi^{(i)}_k
{}~, ~~~i = 1,\ldots,r ~,
\label{4.1a}
\en
we can write this coefficient in terms of the eigenvectors
in the form
\be
\chi^{(\eta)}_k = C_1 \xi^{(1)}_k + \ldots + C_r \xi^{(r)}_k ~,
\label{4.1b}
\en
where the undetermined parameters are exhibited.
 These considerations can be generalized to cases with
several positive eigenvalues. In special situations, also logarithmic
factors ($\lg \la$), and powers thereof,
 may be required \cite{RON,CON}.

For our previously discussed example with Yukawa and quartic coupling, we
have a positive eigenvalue for the solution $\la_{1 +} (\la)$,
which is given by  $\eta = \sqrt{145}/5$.
Consequently, there is a general solution $f_{1+} (\la) = f^0_{1+} + C_+
\la^{\sqrt{145}/5} + \chi^{(3)} \la^3 + \ldots$, with $C_+$ arbitrary,
and the other coefficients being determined.
We see that here the power series solution $f_{1+}(\la) =
f^0_{1+}$ is a stable
solution.  All general solutions also tend to $f^0_{1+}$
for $\la\rightarrow 0$ ,

In the following, we consider briefly a phenomenological application of
general solutions.  For some systems it is not possible to perform a
complete reduction leaving only a single coupling constant, but it may be
sensible to treat a subset of coupling parameters as small perturbations,
while reducing all others.  Suppose we consider the parameters $f_a (\la)
= \la_a (\la) \la^{-1}$,~ $a=1,\ldots,n^\prime$,~ $n^\prime < n$ as small,
and the rest of the couplings, $f_\alpha (\la) =
 \la_\alpha (\la)\la^{-1}$, $\alpha =
n^\prime + 1,\ldots, n$ as large couplings.  If $f_a (\la)\equiv 0$,
we have an undisturbed system involving only the large couplings $f_\alpha$,
and we can look for special solutions as power series in $\la$, with
$f^0_\alpha > 0$, $\la_\alpha = \la f^0_\alpha$.

If we now want to include the small couplings
$f_a (\la)$ as corrections, we look for
solutions of the full set of reduction equations, assuming $f^0_a = 0$
for the small couplings.  Under these circumstances, it is reasonable to
treat $f_\alpha$ as functions of $\la$ and as functionals of the small
couplings $f_a (\la)$. Then we can rewrite the full set of reduction
equations (\ref{3.2}) as partial differential equations for
$f_\alpha (\la,f_a (\la))$ \cite{KMZ,KSZ}:
\be
\beta_\la \la \frac{\partial f_k}{\partial\la} + \sum^{n^\prime}_{a=1}
\frac{\partial f_\alpha}{\partial f_a} (\beta_a (f) - \beta_\la f_a) ~ =
{}~ \beta_\alpha (f) - \beta_\la f_\alpha ~,
\label{4.2}
\en
with solutions $f_\alpha (\la)$, which are power series in $\la$:
\be
f_\alpha (\la) = f^0_\alpha + \sum_{m=0} \la^m \chi^{(m)}_\alpha (f_a
(\la)) ~,
\label{4.3}
\en
The equations (\ref{4.2}) are equivalent to the corresponding
original reduction equations (\ref{3.2}).

As a consequence of the fact that, for $f_a (\la) \equiv 0$, we have the
usual special solution of the subsystem of large couplings, the
coefficients $\chi^{(m)} (f_a)$ can be considered as power
series in $f_a$ \cite{WZZ}.  Inserting
(\ref{4.3}) into (\ref{4.2}), we find that lowest order information is
again sufficient to provide conditions for the uniqueness of the
solutions.  For simplicity, let us make the assumption, which is not
unrealistic for applications, that the matrix $\beta^{-1}_{\la 0}
M(f^0_\alpha)$ is diagonal, and that the eigenvalues $\eta^{(a)} =
\beta^{-1}_{\la 0} M_{aa} (f^0_\alpha)$ are all positive.  With $f^0_a =
0$, the leading behavior of $f_a(\la)$ for $\la\rightarrow 0$ is then
$f_a (\la)\cong C_a\la^{\eta^{(a)}}$.  If this the case, and if we have
an asymptotically free system for $f_a(\la) \equiv 0$, then also the corrected
system is asymptotically free.  Since the coefficients $C_a$ in the
leading terms of $f_a (\la)$ for $\la
\rightarrow 0 $ is arbitrary, we have $n^\prime$
undetermined constants in the reduced solution, which must be fixed by
information about the small couplings in order to get definite values for
the corrections to the undisturbed reductions.

\vskip0.2truein
\centerline{\bf 5.  EXAMPLES}
\vskip0.1truein

As we have mentioned in the introduction, there are many and varied
applications of the reduction method.  Here we can describe only two
examples rather briefly.

Consider an $SU(2) $ gauge theory with one Dirac field and two scalar
fields, all in the adjoint representation.  The interaction part of the
renormalizable theory is then given by
\begin{eqnarray}
{\L}_{int} &=& gauge~ couplings - i\sqrt{\la_1}~
\epsilon^{abc} \overline{\psi}^a (A^b + i\gamma_5 B^b ) \psi^c \cr
               &-& \frac14 \la_2 (A^a A^a + B^a B^a)^2 +
\frac14 \la_3 (A^a A^b + B^a B^b)^2 ~~.
\label{5.1}
\end{eqnarray}
The algebraic reduction equations (\ref{3.6})
have four solutions.  With $\la =
g^2$ as the primary coupling, $g$ being the gauge coupling, these
solutions are power series in $g^2$ for
$g^2 \rightarrow 0 $.  After reparametrization,
they are of the form \cite{OSZ}
\be
(a) ~~~\la_1 = \la_2 = \la_3 = g^2 ~,
\label{5.3}
\en
which corresponds to an $N = 2 $ extended SUSY Yang-Mills theory, and
\be
(b) ~~~\la_1 = g^2,~~ \la_2 = \frac{9}{\sqrt{105}}~ g^2, ~~
\la_3 = \frac{7}{\sqrt{105}}~  g^2 ~,
\label{5.4}
\en
which is a new theory with one coupling and no supersymmetry.
The other two solutions are obtained from (a) and (b) by reversing the
sign of the quartic couplings $\la_2$ and $\la_3$, so that the classical
potential approaches $-\infty $ with increasing scalar fields in almost all
directions.  Hence (a) and (b) are the only acceptable special solutions,
which are power series in $g^2$ with real coefficients.  Note that the Yukawa
coupling $(\la_1)$ is required for the consistency of the reduction.
There are no real solutions of the reduction equations for $\la_1 = 0$.
The two reduced theories (\ref{5.3}) and (\ref{5.4}) can be considered
as {\it minimally} coupled gauge theories with matter fields. As the
original multi-parameter theory, they are asymptotically free.

For the two solutions (\ref{5.3}) and (\ref{5.4}), the eigenvalues of the
$3\times 3$ matrix $\ba^{-1}_{g0} M(f^0) $ are given by
$(-2, ~-3, ~+\frac12)$ for (a), and $(-2, ~-\frac34 (25 \pm
\sqrt{345} )/ \sqrt{105} )$ for (b).
There are no positive integer eigenvalues.
The only positive non-integer eigenvalue
is $\eta = +\frac12 $ in (a).  It gives rise to a general solution
involving $\sqrt{g^2}$, which has one arbitrary parameter.
This solution corresponds to a theory with
hard SUSY breaking, which is consistent with the renormalization group.

Finally, I mention briefly the reduction \cite{KMZ} of the SU(5) minimal SUSY
GUT
\cite{DIM}.
The method of calculations is essentially along the lines we have
discussed in Section
4.  The reduction relates Yukawa couplings and the SU(5) gauge coupling
above the unification scale.  It is required to result in an
asymptotically free theory.  The results are used as boundary conditions
for the conventional application of the
renormalization group for scaling down to the Z--mass.
Of course, a SUSY breaking scale enters into the calculation, but the
results are not very sensitive to it.

I discuss here the gauge-Yukawa unification in the third generation
as an example of the general reduction method. I do not consider the
phenomenology in any detail, nor other approaches to the problem.

The original theory contains three Fermion
families $({\bf {\overline{5}}} + {\bf 10} )$ , two Higgs
supermultiplets $({\bf 5} +{\bf { \overline{5}}} )$  for electroweak symmetry
 breaking, and a ${\bf 24} $ multiplet for the
spontaneous breaking $SU(5) \rightarrow SU(3)\times SU(2)\times U(1)$.
Dimensionfull parameters and family mixing are neglected.  There are six
Yukawa couplings and two Higgs couplings.
With SU(5) indices suppressed, the superpotential is exactly given by
\be
W = \frac12 (g_u {\bf 10}_1 {\bf 10}_1 + g_c {\bf 10}_2 {\bf 10}_2 + g_t
{\bf 10}_3 {\bf 10}_3) H \cr
+ (g_d {\overline{\bf{5}}}_1 {\bf{10}}_1 + g_s {\overline{\bf{5}}}_2
{\bf{10}}_2 + g_b \overline{\bf{5}}_3 {\bf{10}}_3 )\overline{H} \cr
+ \frac{g_l}{3} ({\bf{24}})^3 + g_f \overline{H} {\bf{24}} H ~~,
\label{5.4a}
\en
with $H$ and $\overline{H}$ denoting the  ${\bf {5}}, {\overline
{\bf{5}}}$ Higgs superfields.
Using the one--loop $\beta$--functions for this minimal SU(5) model, we obtain
the reduction equations
\be
\lambda \frac{df_u}{d\lambda} = \frac{27}{5} f_u - 3 f^2_u - \frac43 f_u f_d -
\frac85 f_u f_f ~,\cr
\lambda \frac{df_d}{d\lambda} = \frac{27}{5} f_d - \frac{10}{3} f^2_d -
f_d f_u - \frac85 f_d f_f ~,
\label{5.5}
\en
and corresponding pairs of equations with $f_u,f_d$ replaced by $f_c,f_s$
or $f_t, f_b$ respectively.  The notation is: $\lambda = g^2/4\pi;~~ \lambda_i
= \lambda f_i (\lambda),~~ i=u,d,c,s,t,b,l,f$.  The remaining differential
equations are
\begin{eqnarray}
\lambda \frac{df_l}{d\lambda} &=& g f_l - \frac{21}{5} f^2_l - f_l f_f ~,\cr
\lambda \frac{df_f}{d\lambda} &=& \frac{83}{15} f_f - \frac{53}{15} f^2_f -
f_f f_u - \frac43 f_f f_d - \frac75 f_f f_l ~,
\label{5.6}
\end{eqnarray}
and corresponding equations involving $f_c,f_s$ and $f_t,f_b$.

The reduction is required to be compatible with the mass spectrum of the
first two generations.  This implies that the Yukawa couplings of these
generations should be treated as small perturbations, so that the gauge
-Yukawa reduction is performed only for the third generation.
Accordingly, we require
\be
f^0_u = f^0_d = f^0_c = f^0_s = 0,\cr
f^0_t > 0~,~~~ f^0_b > 0~,
\label{5.7}
\en
and positive eigenvalues $\eta_i$ of the determinant, so that the
corrections preserve the asymptotic freedom of the reduced theory.  As we
have seen in Section 4, the functions
$f_a(\lambda)$ with $f^0_a = 0$ are of the form
$f_a(\lambda) = C_a\lambda^{\eta_a} + \cdots$.

With the conditions described above, the a priori large number of
solution of the equations (\ref{5.5},\ref{5.6}) is reduced to two, which are
given by $\lambda_\alpha = \lambda f^0_\alpha~, ~~~\alpha = t,b,l,f$~, with
\begin{eqnarray}
&(1)&~~ f^0_t = \frac{2533}{2605} = 0.97, \quad f^0_b = \frac{1491}{2605} =
0.57,\quad f^0_l = 0, \quad f^0_f = \frac{560}{521} ~,\cr
&(2)&~~ f^0_t = \frac{89}{65} = 1.37,\quad f^0_b = \frac{63}{65} = 0.97,
\quad f^0_l = 0, \quad f^0_f = 0 ~.
\label{5.8}
\end{eqnarray}
The numbers for the positive eigenvalues $\eta_i$ are given in \cite{KMZ}.

We can now use the solutions (\ref{5.8}) in order to obtain reduction
formulas including small corrections.  The method has been described in
Section 4.  As shown in detail in \cite{KMZ}, on the basis of one--loop
information,  we find that the partial differential equations (\ref{4.2})
have {\it unique} power series solutions of the form (\ref{4.3}), with
coefficients which are again power series in the small couplings.  It
turns out that the contributions from the Yukawa couplings of the first
two generations are negligible.  Hence we obtain the solutions
\begin{eqnarray}
(1)~~ f_\alpha &=& f^0_\alpha + \chi^{(1)}_{l\alpha} f_l +
\chi^{(2)}_{l\alpha} f^2_l + \cdots, \cr
\alpha &=& t,b,f~, \cr
(2)~~ f_\alpha &=& f^0_\alpha + \chi^{(1)}_{l\alpha} f_l +
\chi^{(11)}_{l f\alpha} f_l f_f + \chi^{(2)}_{l\alpha} f^2_l
+ \chi^{(2)}_{f\alpha} f^2_f + \cdots,    \cr
\alpha &=& t,b.
\label{5.9}
\end{eqnarray}
The coefficients $\chi$ are determined, and we have $f_l = C_l\lambda
^{\eta_l},~~ f_f = C_f\lambda^{\eta_f}$ with positive eigenvalues $\eta$, while
$C_l$ and $C_f$ are to be fixed by input data. Actually, the two special
solutions listed above are boundaries of the same, asymptotically free
general solution.

The solutions (\ref{5.8}) for the reduction with corrections of the minimal
SU(5) model are valid at the unification scale ($\sim 10^{25}$ GeV).  As
mentioned before, they can be used as input for the usual renormalization
group equations of the minimal supersymmetric standard model,
which are set up to scale the reduced theory down to
energies of the order of $M_Z$.  A SUSY breaking mechanism must be
introduced, so that the superpartners are unobservable at lower energies.
There
are other requirements, like the suppression of proton decay, which must
be taken into account.  With the input data
$$
\sin^2 \theta_W(M_Z) =0.2315~,~
\alpha_{\rm em}^{-1}(M_Z)=128.09~,$$ $$
{}~m_{\tau} = 1.777 ~\mbox{GeV}~,~
M_Z=91.188 ~\mbox{GeV}~,
$$
and SUSY breaking scales of 300 or 500 GeV, the main results obtained by
Kubo, Mondrag\'{o}n and Zupanos
\footnote{I am very grateful to Jisuke Kubo for kindly providing me
with the data from the latest recalculation.
These results are essentially the same as those obtained
in March 1994  \cite{KPR,KMZ}. }
are
$$
(1) ~~m_t=179.0~\mbox{GeV}~,~m_b=5.5~\mbox{GeV}~,
{}~\alpha_S(M_Z)=0.123~$$
$$
(2) ~~m_t=182.3~\mbox{GeV}~,~m_b=5.4~\mbox{GeV}~,
{}~\alpha_S(M_Z)=0.123~$$
for $M_{\rm SUSY}=300$ GeV, and
$$
(1) ~~m_t=179.0~\mbox{GeV}~,~m_b=5.5~\mbox{GeV}~,
{}~\alpha_S(M_Z)=0.122~$$
$$
(2) ~~m_t=182.6~\mbox{GeV}~,~m_b=5.3~\mbox{GeV}~,
{}~\alpha_S(M_Z)=0.122~$$
for $M_{\rm SUSY}=500$ GeV.

Here the masses are the pole masses, and  $\alpha_S(M_Z)$
is defined in the $\overline{\rm MS}$ scheme with
five flavors. These results  are obtained
in an asymptotically free, renormalizable, reduced theory.
Reduction calculation have also been performed starting from different initial
models \cite{KUB}.  The results are quite similar.

As I have mentioned before, this talk is mainly
concerned with  a description of the reduction method,
and not with phenomenology.
But in order to assess the results for the top-bottom hierarchy, it is
relevant to consider an aspect of the renormalization group equations
for scaling the Yukawa couplings from the GUT scale to the scale of the
$Z$-mass. This scaling is done within the framework of the minimal
supersymmetric
standard model. It turns out that, for sufficiently large values of the
top-Yukawa coupling $\lambda_t $ at the
unification scale, the results obtained for $m_t (M_Z)$ are rather insensitive
to the precise value of this coupling.
Consequently, calculations of the coupling $\lambda_t$ can be tested more
sensitively if the experimental top-quark mass is at least slightly smaller
than
the value corresponding to the plateau of the curve $m_t(M_Z)$ versus
$\lambda_t$.
\footnote {In the literature, this value of the top-quark mass is usually
called
the ``quasi infrared fixed point'' solution \cite{CHR}. }

The reduction method has previously been
applied to the standard theory for
electroweak and strong interactions \cite{KSZ}.
  Extensive calculations, including
all corrections, give top masses (about $100~\mbox{GeV}$) well below the
experimental value.
This is an important result, because it indicates that more fields are
needed, which are provided for in the SUSY models described above.

\vskip0.7truein
\centerline{\bf ACKNOWLEDGMENTS}
\vskip0.2truein

For conversations and remarks.  I am indebted to K. Sibold and
W. Zimmermann. I am particularly grateful to Jisuke Kubo for many
helpful communications.
It is a pleasure to thank Wolfhart Zimmermann, and the
theory group of the Max Planck Institut f\"ur Physik, Werner Heisenberg
Institut, for their kind hospitality in M\"unchen.

This work has been supported in part by the National Science Foundation,
grant PHY 91-23780.

\end{document}